\documentclass[12pt,a4paper]{article}
\usepackage{amsmath,amssymb,amsfonts}
\usepackage{graphicx,a4wide}
\usepackage[top=3.0cm,bottom=2cm,left=1.5cm,right=1.5cm]{geometry}

\newcommand{\be}{\begin{equation}}
\newcommand{\ee}{\end{equation}}
\newcommand{\bea}{\begin{eqnarray}}
\newcommand{\eea}{\end{eqnarray}}

\newcommand{\nn}{\nonumber}

\newcommand{\al}{\alpha}
\newcommand{\gm}{\gamma}
\newcommand{\Gm}{\Gamma}
\newcommand{\ep}{\epsilon}
\newcommand{\de}{\delta}
\newcommand{\De}{\Delta}
\newcommand{\om}{\omega}

\newcommand{\tht}{\theta}

\newcommand{\lm}{\lambda}
\newcommand{\Lm}{\Lambda}
\newcommand{\sg}{\sigma}

\newcommand{\oD}{\overline{D}}

\newcommand{\oq}{\overline{q}^{\,2}}
\newcommand{\ov}{\overline}

\newcommand{\op}{\overline\Pi}
\newcommand{\iop}{{\rm Im}\,\overline\Pi}
\newcommand{\od}{\overline D}  
\newcommand{\wt}{\widetilde}
\newcommand{\tom}{\widetilde{\om}}   

\newcommand{\vk}{\vec k}
\newcommand{\vp}{\vec p}
\newcommand{\vq}{\vec q}
\newcommand{\vx}{\vec x}
\newcommand{\la}{\langle}
\newcommand{\ra}{\rangle}
\newcommand{\bL}{\boldsymbol{\Lambda}}

\newcommand{\bD}{\boldsymbol{D}}
\newcommand{\bU}{\boldsymbol{U}}

\newcommand{\ks}{k \!\!\! /}
\newcommand{\qs}{q \!\!\! /}

\newcommand{\rw}{\rightarrow}
\newcommand{\mn}{{\mu\nu}}
\newcommand{\del}{\partial}

\newcommand{\F}{F_\pi}

\def\bet{\beta}
\def\negs{\!\!\!\!\!\!\!\!}
\def\no{\nonumber}
\def\th{\theta}

\begin{document}

\setcounter{page}{1}

\title{
Analysis of $\om$ self-energy at finite temperature and density in the real-time
formalism
}
\author{
Sabyasachi Ghosh and Sourav Sarkar
}
\maketitle
\begin{center}
\it{Theoretical Physics Division, Variable Energy Cyclotron Centre,\\
1/AF, Bidhannagar, 
Kolkata 700064, India}
\end{center}

\begin{abstract}
Using the real time formalism of field theory at finite temperature and density
we have evaluated the in-medium $\om$ self-energy from baryon and meson loops. 
We have analyzed in detail the discontinuities across the branch cuts
of the self-energy function and obtained the imaginary part from the 
non-vanishing contributions in the cut regions.
An extensive set of resonances have been 
considered in the baryon loops. Adding the meson loop contribution
we obtain the full modified spectral function of the $\om$ meson in a thermal gas of mesons, 
baryons and anti-baryons in equilibrium for 
several values of temperature and baryon chemical potential.

\end{abstract}

%\pacs{11.10.Wx}

%\keywords{}
\maketitle

\section{Introduction}

The in-medium properties of vector mesons has been a much discussed 
topic~\cite{Annals,CBMBook,Hayano,Leupold,SS_npa}. One 
of the reasons for this is the possible connection with the (partial) restoration
of the spontaneously broken chiral symmetry of QCD at high temperature and/or
density.  The search for this phenomenon is one of the driving motivations of
relativistic heavy ion collision experiments around the world. The invariant mass
spectra of lepton pairs is the most promising observable. The fact that in the
low mass region the emission rate of dileptons is proportional to the spectral
function of low lying vector mesons is the main interest behind their study.

A large volume of literature is dedicated to the study of vector
mesons in the medium, the bulk of which concerns the $\rho$ meson. 
Theoretical activities regarding the $\om$ meson have also
received some attention~\cite{Leu_38,
Oset_4,Oset_7,Oset_8,Oset_10,Oset_12,Oset_13,Lutz,Weise,
Muelich,Leu_9,Leu_50,Leu_51,Weise_Wass,Leu_39,Leu_42,Leu_46,Leu_43,Murat,
Sneider,Eletsky,Leu_48,Rapp_prc,Roy,Alam,Leu_49}.
Though the low density expansion has been 
used in most cases~\cite{Lutz,Weise,Muelich} the approaches differ widely in 
their methods resulting in a large variation in the results concerning the mass and
width. Consequently positive~\cite{Lutz,Leu_9,Leu_50,Leu_51}, 
negative~\cite{Weise,Weise_Wass,Leu_39,Leu_42,Leu_46,Leu_43} and
zero~\cite{Muelich} shift of the peak position have
been proposed. Using a different approach~\cite{Murat}, 
analysis of the $\gamma A\to \pi^0\gamma X$ reaction 
resulted in a large width and no mass shift
at nuclear matter density. On the experimental front
also the situation is far
from settled~\cite{Hayano} with different groups reporting a reduction in 
mass~\cite{KeK} and increase in width~\cite{Taps} in $pA$ and $\gm A$ 
collisions respectively. 

Barring a small number, discussions on the in-medium spectral properties 
of the $\om$ meson have been mostly carried out in cold nuclear matter.
Finite temperature calculations at vanishing baryon density have
been done in~\cite{Sneider} showing a large increase in width due to $\om\to 3\pi$
and $\om\pi\to\pi\pi$ processes. Baryon induced effects on the $\om$ spectral
function at finite temperature have been treated within a virial approach
in~\cite{Eletsky} where the self-energy was obtained in terms of vector
meson-pion and nucleon scattering amplitudes constructed using resonance
dominance at low energies and Regge approach at higher energies. This estimate was improved upon in~\cite{Leu_48} using better 
resonance data in the $\om N$ channel obtaining a large enhancement in the width
without any significant mass shift. 
Similar conclusions were made in~\cite{Rapp_prc} where in addition to 
contributions coming from scattering with mesons, resonance-hole contributions 
have been included in the self-energy. The $\om$ meson has also been studied 
at finite temperature and density within the Walecka model~\cite{Roy,Alam}
obtaining a decrease in mass and an increase in width. Similar conclusions
were made in~\cite{Leu_49} using a many body approach.

It is thus evident that the modification of the mass and width of the $\omega$
meson embedded in hot and/or dense hadronic matter is yet to reach
a definite conclusion. It is essential to realize in this connection that the vector currents
with finite three-momentum can experience complex interactions with thermal
excitations as a result of which the vector spectral function could exhibit new 
structures
apart from the familiar shift of the peak and broadening of the width. The
entire spectral shape is thus of interest. This requires a detailed
field theoretic analysis of the spectral function at finite temperature,
baryon density and three-momentum which is necessary to understand the
signatures of chiral symmetry restoration from the
analysis of the low mass dilepton spectra from heavy ion collisions. 
This is the principal motivation of the present study.  

Recently, the $\rho$ spectral function was evaluated at finite
temperature and baryon density~\cite{GSM_EPJC,GS_NPA} in which the sources
 modifying the free
propagation of the $\rho$ was obtained in a unified way from the branch cuts of the
self-energy function. It was shown that the spectral strength at lower invariant
masses was significantly enhanced due to contributions coming from the Landau cut
which appears only in the medium and provides the effect of collisions of the
$\rho$ with the particles in the bath. Here we extend this analysis for the
$\om$ meson. For the baryonic contribution we have considered an extensive number
of spin one-half and three-half resonances in the one-loop diagrams. These 
diagrams have been evaluated using the real time formulation of thermal field
theory using the full relativistic baryon propagators in which baryons and
anti-baryons appear on an equal footing. As a result, distant singularities
coming from the unitary cut involving heavy baryons in the loop, which are
neglected in certain approaches, are automatically included. They are found
to contribute appreciably to the real part of the self-energy~\cite{GSS_PRC_breif}.

The article is organized as follows. In the following section we define the 
spectral function of the $\om$ in terms of the diagonal component of the 
thermal self-energy matrix which appears in the real time formalism. 
In section 3 we evaluate the 11-component of the $\om$ self-energy matrix
for one-loop diagrams containing mesons and baryons and analyse their
singularity structure. We have presented results of 
numerical evaluation of the real and imaginary parts of the self-energy
as well as the spectral function of $\om$ in section 4. We summarize
and conclude in section 5. A self-contained discussion of the analytic structure
of the vector meson propagator and its spectral
representation is provided in Appendix-A. Details of the interaction Lagrangian
and terms appearing in the self-energy have been provided in Appendix-B. 
In Appendix-C it is shown how the real part obtained by a direct evaluation
can be equivalently put in a dispersion integral form.

\section{The full $\om$ propagator in the medium}

We begin with the complete $\om$-propagator in vacuum given by the Dyson
equation,
\be
D_{\mn}(q)=D_{\mn}^{(0)}(q)- D_{\mu\lm}^{(0)}(q)\Pi^{\lm\sg}(q) D_{\sg\nu}(q)
\label{dyson_G_vac}
\ee
where 
\be
D^{(0)}_{\mn}(q)=\left(-g_{\mn}+
\frac{q_\mu q_\nu}{m_\om^2}\right)\De(q)~;~~~~\De(q)=\frac{-1}{q^2-m_\om^2+i\ep}
\label{freeD}
\ee
is the free propagator.

In the real-time formulation of thermal field theory all two-point functions
assume a $2\times 2$ matrix structure on account of the contour in the complex  
time plane (see Appendix-A). As a result the Dyson equation for the full
propagator in the medium is given by the matrix equation
\be
D_{\mn}^{ab}(q)=D_{\mn}^{(0)ab}(q)- D_{\mu\lm}^{(0)ac}(q)\Pi^{cd,\lm\sg}(q)
D_{\sg\nu}^{db}(q)
\label{dyson_G_matrix}
\ee
where $a,b,c,d$ are thermal indices and take values 1 and 2. The free thermal
propagator is given by
\be
D_{\mn}^{(0)ab}(q)=\left(-g_{\mn}+\frac{q_\mu q_\nu}{m_\om^2}\right)D^{ab}(q)
\ee
where the matrix $D^{ab}$ has the components 
\bea
&&D^{11}=-(D^{22})^*=\De(q)+2\pi in\de(q^2-m_\om^2)\nonumber\\
&&D^{12}=D^{21}=2\pi i\sqrt{n(1+n)}\de(q^2-m_\om^2)
\label{Dlam}
\eea
$n$ being the Bose distribution function. The thermal indices can however be
removed by diagonalization resulting in the equation
\be
\od_{\mn}(q)=\od_{\mn}^{(0)}(q)-\od_{\mu\lm}^{(0)}(q)\op^{\lm\sg}(q)
\od_{\sg\nu}(q)
\label{dyson_G_diag}
\ee
where the quantities with bars denote the corresponding diagonal components
and in particular $\od_\mn^{(0)}(q)=D_\mn^{(0)}(q)$ as given in (\ref{freeD}).
Decomposing the self-energy function $\op_{\mn}(q)$ into transverse and
longitudinal components using the projection operators $P_\mn$ and $Q_\mn$
respectively, eq.~(\ref{dyson_G_diag}) can now be solved analogously
as in vacuum to get 
\be
\od_\mn(q)=\frac{-P_\mn}{q^2-m_\om^2-\op_t(q)}+
\frac{-Q_\mn/q^2}{q^2-m_\om^2-q^2\op_l(q)}+\frac{q_\mu q_\nu}{q^2 m_\om^2}
\label{Dmn}
\ee
where~\cite{GS_NPA}
\be
P_{\mn}=-g_{\mn}+\frac{q_\mu q_\nu}{q^2}-\frac{q^2}{\oq}\wt u_\mu \wt
u_\nu,~\wt u_\mu=u_\mu-(u\cdot q)q_{\mu}/q^2
\label{defP+Q}
\ee
and
\be
Q_{\mn}=\frac{(q^2)^2}{\oq}\wt u_\mu \wt
u_\nu,~\oq=(u\cdot q)^2-q^2~.
\ee
$u_\mu$ being the four-velocity of the thermal bath.
The components of the self-energy function are defined as
\be
\op_t=-\frac{1}{2}(\op_\mu^\mu +\frac{q^2}{\bar q^2}\op_{00}),~~~~
\op_l=\frac{1}{\bar q^2}\op_{00} , ~~~\op_{00}\equiv u^\mu u^\nu \op_{\mn}
\label{pitpil}
\ee
which can be obtained from 
the 11-component of the in-medium self-energy matrix using
\bea
{\rm Re}\,\op_{\mn}&=&{\rm Re}\,\Pi_{\mn}^{11}
\nonumber\\
\iop_{\mn}&=&\epsilon(q_0)\tanh(\beta q_0/2){\rm Im}\,\Pi_{\mn}^{11}~.
\label{def_diag}
\eea

A few comments on renormalization are in order here. Note that after 
performing the mass and field renormalizations  
the $\om$ propagator actually has the form
\be
\od_\mn(q)=\frac{-P_\mn}{q^2-m_\om^2-\widetilde\Pi_{vac}(q^2)-\op_t(q)}+
\frac{-Q_\mn/q^2}{q^2-m_\om^2-\widetilde\Pi_{vac}(q^2)-q^2\op_l(q)}+\frac{q_\mu q_\nu}{q^2 m_\om^2}
\ee
where $\widetilde\Pi_{vac}(q^2)$ is the zero-temperature self-energy that
remains after subtracting out the first two Taylor terms,
\be
\widetilde\Pi_{vac}(q^2)=\Pi_{vac}(q^2)-\Pi_{vac}(m_\omega^2)-(q^2-m_\omega^2)\Pi_{vac}^{\,\prime}(m_\omega^2)
\ee
and $\op_{t/l}(q)$ is the temperature dependent part. 
Thus $\widetilde\Pi_{vac}(q^2)\sim O[(q^2-m_\om^2)^2]$. As we are working around
$q^2=m_\om^2$, $\widetilde\Pi_{vac}(q^2)$ is rather small. 
We thus include only the thermal parts $\op_{t/l}(q)$ in
eq.~(\ref{Dmn}) which are free from divergences up to one-loop order. 
 
We also note in this context that we are dealing with a theory which is not
Dyson-renormalizable. Such effective theories are valid in the low energy region
and there is no way of imposing any asymptotic constraint.

\section{Analytical structure of $\omega$ meson self-energy}
\begin{center}
\begin{figure}
\includegraphics[scale=1.3]{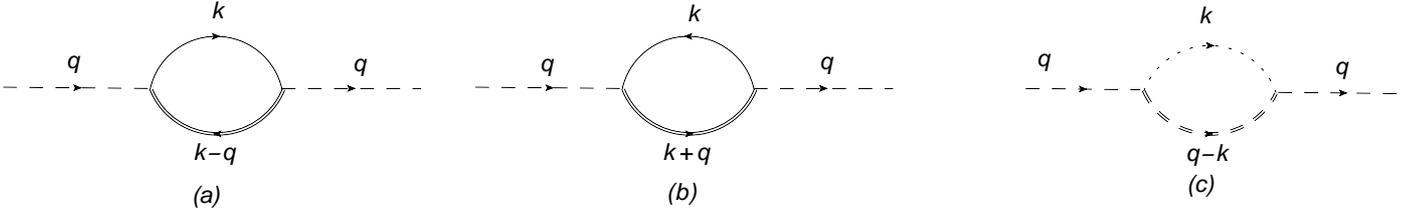}
\caption{One-loop $\om$ self-energy diagrams 
with baryons (a) and (b) where single and double lines represent nucleon $(N)$ 
and resonances $(B)$ respectively. Diagram (c) indicates meson loop where 
dashed, dotted and double dashed lines stand
for $\om$, $\pi$ and $\rho$ respectively.}
\label{loop_BB}
\end{figure} 
\end{center}

In this section we will
evaluate one-loop diagrams of the type shown in Fig.~\ref{loop_BB} with mesons 
and baryons in the internal
lines in order to obtain the leading temperature and density effect modifying
the free propagation of the $\om$ meson. The two cases are separately discussed
below.

\subsection{Baryon Loops}

The internal lines in these loops contain a nucleon $N$ and a baryon $B$ which 
represents 
several spin one-half and three-half $4-$star
resonances. Here $B$ stands for the $N^*(1440)$
$N^*(1520)$, $N^*(1535)$, $N^*(1650)$, $N^*(1720)$ resonances
as well as the $N(940)$. 
For spin 1/2 resonances in the loop, the expression for $11$-component of the $\om$ self-energy 
in medium is given by
\be
\Pi^{11,\mn}(q)=i I_F\sum_{a=-1,+1}
\int \frac{d^4k}{(2\pi)^4} Tr[\Gm^{\mu}S^{11}(k,m_N)\Gm^{\nu}S^{11}(k+a q, m_B)] 
\label{spin1/2}
\ee
where $S^{11}$ is the 11-component of the thermal propagator for fermions 
in the real-time formalism which is defined as
$S^{11}(k,m)=(\ks+m)E^{11}(k,m)$  with
\bea
E^{11}(k,m)&=&\frac{-1}{k^2-m^2+i\ep}-2\pi i
N(k_0)\delta(k^2-m^2);~~~~~~~~N(k_0)=n_+(\omega)\theta(k_0)+
n_-(\omega)\theta(-k_0)\nonumber\\
\nonumber\\
&=&-\frac{1}{2\om}\left(\frac{1-n_+}{k_0-\om+i\ep}+
\frac{n_+}{k_0-\om-i\ep}-\frac{1-n_-}{k_0+\om-i\ep}
-\frac{n_-}{k_0+\om+i\ep}\right)\,
\label{de11}
\eea
The function $n_\pm(\omega)=\displaystyle\frac{1}{e^{\beta(\omega \mp \mu_B)}+1}$
is the Fermi distribution for energy $\om=\sqrt{\vk^2+m^2}$ where the $\pm$ 
sign in the subscript refers to baryons and 
anti-baryons respectively  
and $\mu_B$ is the baryonic chemical potential which is 
taken to be equal for all baryons considered here. 
In the second line of eq.~(\ref{de11}), the first and the second terms 
are associated with the propagation of baryons
above the Fermi sea and holes in the Fermi sea respectively while
the third and
fourth terms represent the corresponding situation for
anti-baryons~\cite{Vol_16}. 
The full relativistic baryon propagator thus treats the
baryons and anti-baryons on an equal footing and the additional singularities 
which are not generally considered in usual approaches
are automatically included. 
The two values of $a$ in (\ref{spin1/2}) correspond to the direct and crossed diagrams shown in Fig.~\ref{loop_BB}
(a) and (b) respectively which are obtainable from one another by changing the 
sign of the external momentum $q$.

The corresponding expression for the case of loop graphs with spin 3/2 resonances is given by
\be
\Pi^{11,\mn}(q)=i I_F \sum_{a=-1,+1}\int \frac{d^4k}{(2\pi)^4} 
Tr[\Gm^{\mu\alpha}S^{11}(k,m_N)\Gm^{\nu\beta}S^{11}_{\beta\alpha}(k+a q, m_B)]
\label{spin3/2}
\ee 
in which the spin-3/2 (Rarita-Schwinger) propagator is given by
\be 
S^{11}_{\mn}(k,m)=(\ks + m)K_{\mn}(k,m) E^{11}(k,m)~.
\ee
The factor
\[K_{\mn}(k,m)= -g_{\mn}+ \frac{2}{3m^2}k_\mu k_\nu + \frac{1}{3}\gamma_\mu\gamma_\nu 
+ \frac{1}{3m}(\gamma_\mu k_\nu - \gamma_\nu k_\mu )
\]
in the numerator ensures that there is no on-shell propagation of non-physical
degrees of freedom which arise in the vector-spinor representation.

Obtaining the vertex factors $\Gm^{\mu}$ and $\Gm^{\mu\alpha}$ 
from the interaction Lagrangians given in Appendix-B,
both (\ref{spin1/2}) and (\ref{spin3/2}) can
be expressed as
\be
\Pi^{11,\mn}(q)=i I_F\sum_{a=-1,+1}\int \frac{d^4k}{(2\pi)^4} L^{\mn}(k,q)
E^{11}(k,m_N) E^{11}(k+a q,m_B)
\label{Pimn-Lmn}
\ee
where the isospin factor $I_F$ is 2 for all the loops.  
The factor $L_{\mn}(k,q)$ consists of a trace over Dirac matrices appearing 
in the two fermion propagators along with
their associated tensor structure coming
from the $\om NB$ vertex. Their explicit forms obtained after a complete
evaluation are given in Appendix-B. As expected, the $L_{\mn}$ (and consequently
the self-energy $\Pi_{\mn}$ for all the diagrams) turn out to be four dimensionally transverse.

Let us first discuss diagram (a) for which $a=-1$.
The diagonal element of the in-medium self energy
defined by eq.~(\ref{def_diag}) 
after integration over $k^0$ can be written as 
\bea
{\ov\Pi}^{\mn}(q)&=&\int\frac{d^3k}{(2\pi)^3}\frac{1}{4\om_N\om_B}\left[\frac{(1-n^N_+)L^{\mn}_1-n^B_-L^{\mn}_3}{q_0 -\om_N-\om_B+i\eta\ep(q_0)}
+\frac{n^N_+L^{\mn}_1-n^B_+L^{\mn}_4}{q_0-\om_N+\om_B+i\eta\ep(q_0)} 
\right.\nn\\
&&+ \left.\frac{-n^N_-L^{\mn}_2 +n^B_-L^{\mn}_3}{q_0 +\om_N-\om_B+i\eta\ep(q_0)} 
+\frac{n^N_-L^{\mn}_2 +(-1+n^B_+)L^{\mn}_4}{q_0 +\om_N+\om_B+i\eta\ep(q_0)}\right]
\label{Pi_a}
\eea
where $n^N\equiv n(\omega_N)$ with $\omega_N=\sqrt{\vk^2+m_{N}^2}$, $n^B\equiv 
n(\omega_B)$ with $\omega_B=\sqrt{(\vk-\vq)^2+m_{B}^2}$
and $L^{\mn}_i,i=1,..4$ denote the values of $L^{\mn}(k_0)$ for
$k_0=\om_N,-\om_N,q_0-\om_B,q_0+\om_B$ respectively. 
The imaginary part of (\ref{Pi_a}) 
is easily obtained and is given by
\bea
{\rm Im}\op^{\mn}&=&-\ep(q_0)\pi\int\frac{d^3k}{(2\pi)^3}\frac{1}{4\om_N\om_B}[L^{\mn}_1\{(1-n^N_+ - n^B_-)\delta(q_0-\om_N-\om_B)
\nn\\
&&+(n^N_+ -n^B_+ )\delta(q_0-\om_N+\om_B)\}+L^{\mn}_2\{(-n^N_- + n^B_-)\delta(q_0+\om_N-\om_B)
\nn\\
&&~~~~~~~~~~~~~+(-1+n^N_- + n^B_+)\delta(q_0+\om_N+\om_B)\}]
\label{ImPi_a}
\eea
in which the factors $L^{\mn}_{3,4}$ have transformed into $L^{\mn}_{1}$ or
$L^{\mn}_{2}$ on use of the associated $\de$-functions. 
The four $\de$-functions give rise to cuts in the self-energy function on the real
axis in the complex energy plane and define the different kinematic domains 
where the imaginary part is non-zero. The first and fourth terms contribute
for $q^2>(m_B+m_N)^2$ and define the unitary cut and the second and third terms
which are non-zero for $q^2<(m_B-m_N)^2$ define the Landau cut.
For a detailed discussion of the cuts in the complex $q_0$ plane
see~\cite{GSM_EPJC}. While the
unitary cut is always present, the Landau cut appears only in the medium.
In the case of the baryon loops only the Landau cut contribution is
relevant to the imaginary part, the threshold for the unitary cuts being 
distant from the $\om$ pole.
It is easy to see that the imaginary part (coming from the Landau cuts) 
arises due to various scattering processes in which the $\om$ 
is gained or lost during its propagation in the medium.
For example, rearranging the statistical weights in the second term of 
eq.~(\ref{ImPi_a}) as 
$(n_+^N-n_+^B)=n_+^N(1-n_+^B)-n_+^B(1-n_+^N)$, this term
can be interpreted as the contribution to the imaginary part due to the
disappearance of the $\omega$ by absorbing a nucleon from the heat bath to
produce a higher mass baryon with 
the Pauli blocked probability $(1-n_+^B)$ minus
the reverse process where it is produced along with the nucleon (which now suffers
Pauli blocking) in the decay of the resonance $B$.
The third term of eq.~(\ref{ImPi_a}) can be analogously interpreted in terms of
scattering with the anti-baryons.

The cut-structure for the second diagram (b) in Fig.~\ref{loop_BB} can also be
analysed in the same way. Restricting to terms contributing to the
physically relevant kinematic region $q^2,\ q_0>0$, the total contribution from the baryon
loops is given by
\bea
\iop^{\mn} (q_0,\vq)&=&\frac{-\ep(q_0)}{16\pi|\vq|}\int_{\tom^{+}_{N}}^{\tom^{-}_{N}} 
d\tom_N [L^{\mn}_1(a=+1)\{-n_+(\tom_N)+n_+(\tom_B=q_0+\tom_N)\}
\nn\\
&&~~~~~~~~~~~~~~+L^{\mn}_2(a=-1)\{-n_-(\tom_N)+n_-(\tom_B=q_0+\tom_N)\}]
\eea
where $\tom^{\pm}_{N}=\frac{S^2_{N}}{2q^2}(-q^0 \pm |\vq| W_{N})$
with $W_{N}=\sqrt{1-\frac{4q^2m_N^2}{S^4_N}}$,
$S^2_{N}=q^2-m_B^2+m_N^2$.

The real part of the self-energy can be obtained by evaluating the 
principal value integrals in eq.~(\ref{Pi_a}) which remain after the 
imaginary parts are removed. 
The terms within square brackets denoted by unity indicate the vacuum
contribution to the real part of the self-energy. After mass and field
renormalization a finite piece of $O(q^2-m_\om^2)^2$ is ignored.
%These divergent
%contributions from the baryonic loops as well as the mesonic loop discussed 
%later are approximately taken into account by assuming that they
%renormalize the $\omega$ mass to its physical value. 
The terms with the Fermi
distribution functions denote the medium contributions to the real part
of the self-energy. These are finite owing to the natural cut-off provided by
the thermal distribution functions. 
Note that at a given value of $q$ the real part 
receives contribution from all the four terms unlike the imaginary part
where the contribution depends on the associated $\de$-function. 
Note that the real and imaginary parts of the self-energy
obtained above can be related by a dispersion integral.
This is shown in Appendix-C. 

The baryon resonances $B$ have so far been considered in the narrow width 
approximation.  
For a realistic treatment it is necessary to include the width of the 
resonances.
For this, we follow the procedure outlined 
e.g. in~\cite{Gonzalez,SarkarNPA}
of convoluting the self energy calculated in the narrow width 
approximation with the spectral function of the baryons as done
for the $\rho$ meson in~\cite{GS_NPA}. 
This approach has the advantage that the analytic structure of the self energy 
discussed above is not disturbed (see Appendix-C). 
\be
\op_B(q,m_B)= \frac{1}{N_B}\int^{m_B+2\Gm_B}_{m_B-2\Gm_B}dM\frac{1}{\pi} 
{\rm Im} \left[\frac{1}{M-m_B + \frac{i}{2}\Gm_B(M)} \right] \op_B(q,M) 
\ee
with $N_B=\displaystyle\int^{m_B+2\Gm_B}_{m_B-2\Gm_B}dM\frac{1}{\pi} {\rm Im} 
\left[\frac{1}{M-m_B + \frac{i}{2}\Gm_B(M)} \right]$ and 
$\Gm_B(M)=\Gm_{B\rightarrow N\pi} (M) + \Gm_{B\rightarrow N\rho}
(M)$; $M=\sqrt{q^2}$.

\subsection{Meson loop}

The 11-component of $\om$ meson self energy for the 
$\rho\pi$ loop is given by
\be
\Pi^{11,\mn}_{(\rho\pi)}(q,m_{\rho})=i 
\int \frac{d^4k}{(2\pi)^4}L^{\mn}(k,q) D^{11}(k,m_{\pi})D^{11}(q-k, m_{\rho})
\label{self_meso}
\ee
where $D^{11}$ is the 11-component of the scalar propagator defined above.
As before, the tensor structure associated
with the two vertices and the vector propagator are included in $L^{\mn}$, the
details of which are given in Appendix-B. 
The diagonal element 
of the thermal self-energy matrix is obtained as
\bea
{\ov\Pi}_{(\rho\pi)}^{\mn}(q,m_{\rho})&=&\int\frac{d^3k}{(2\pi)^3}\frac{1}
{4\om_{\pi}\om_{\rho}}\left[\frac{(1+n^{\pi})L^{\mn}_1+n^{\rho}L^{\mn}_3}
{q_0 -\om_{\pi}-\om_{\rho}+i\eta\ep(q_0)}
+\frac{-n^{\pi}L^{\mn}_1+n^{\rho}L^{\mn}_4}
{q_0-\om_{\pi}+\om_{\rho}+i\eta\ep(q_0)} 
\right.\nn\\
&&+\left. \frac{n^{\pi}L^{\mn}_2 -n^{\rho}L^{\mn}_3}{q_0 +\om_{\pi}-\om_{\rho}+i\eta\ep(q_0)} 
+\frac{-n^{\pi}L^{\mn}_2 -(1+n^{\rho})L^{\mn}_4}
{q_0 +\om_{\pi}+\om_{\rho}+i\eta\ep(q_0)}\right]
\label{MM_rho}
\eea
where $n^\pi=n(\om_\pi)$ with  $\om_\pi=\sqrt{\vk^2+m_\pi^2}$ and
$n^\rho=n(\om_\rho)$ with  $\om_\rho=\sqrt{(\vq-\vk)^2+m_\rho^2}$. Here,
unlike the baryon loops 
both the Landau and unitary cuts are relevant for the imaginary part 
in the kinematic domain of our interest. These are respectively given by
\be
{\rm Im}{\ov\Pi}^{\mn}_{(\rho\pi)}=-\frac{\ep(q_0)}{16\pi |\vq|}\int_{{\tom^+_{\pi}}}^{{\tom^-_{\pi}}} d\tom_{\pi} L^{\mn}_2
\{n(\tom_{\pi})-n(\tom_{\rho}=q_0+\tom_{\pi})\}
\label{im_L}
\ee
and
\be
{\rm Im}{\ov\Pi}^{\mn}_{(\rho\pi)}=-\frac{\ep(q_0)}{16\pi |\vq|}\int_{\om^-_{\pi}}^{\om^+_{\pi}} d\om L^{\mn}_1
\{1+n(\om_{\pi})+n(q_0-\om_{\pi})\}
\label{im_U}
\ee
where the integration limits
$\om^{\pm}_{\pi}=\frac{S^2_{\pi}}{2q^2}(q_0\pm |\vq|W_{\pi})$, 
${\tom^{\pm}_{\pi}}=\frac{S^2_{\pi}}{2q^2}(-q_0\pm |\vq|W_{\pi})$
with $W_{\pi}=\sqrt{1 -\frac{4q^2 m_\pi^2}{S^4_{\pi}}}$ 
and $S^2_{\pi}=q^2-m^2_{\rho}+m^2_{\pi}$.

The real part of the self-energy can be easily read off from (\ref{MM_rho}) 
in terms of principal value integrals and we do not write them here.

The $\om$ self-energy due to its coupling to $3\pi$ states can be estimated
by folding the $\rho\pi$ contribution 
with the $\rho$ spectral function $A_\rho$ as in~\cite{Muelich} getting
\be
{\ov\Pi}^{\mn}_M(q)=\frac{1}{N_{\rho}}\int^{(q-m_{\pi})^2}_{4m^2_{\pi}} 
dM^2[{\ov\Pi}^{\mn}_{(\rho\pi)}(q,M)]A_{\rho}(M)
\label{3pi_med}
\ee
where $N_{\rho}=\int^{(q-m_{\pi})^2}_{4m^2_{\pi}} dM^2 A_{\rho}(M)$
and $A_{\rho}$ is the $\rho$ spectral function. 

Following~\cite{Weldon,GSM_EPJC}, the Landau cut contribution from the $\rho\pi$ loop 
can be interpreted as 
the probability of occurrence of processes like $\om\pi\to\rho$ and 
$\om\rho\to\pi$ which are responsible for the loss of $\om$ mesons in the
medium minus the reverse processes which lead to a gain. Similarly, the 
unitary cut contribution accounts for processes like $\om\to\rho\pi$ and
its reverse. As  a consequence of folding with the $\rho$ spectral function
containing its $2\pi$ decay width, all possible scatterings 
like $\om\pi\to\pi\pi$, $\om\pi\pi\to\pi$ etc. as well as the decay $\om\to
3\pi$, proceeding through $\rho$-exchange are effectively accounted 
for in the imaginary part. 
%As we will see in
%Fig.~\ref{M_Btot} in the next section, the sharp cut-off defining the end of
%the Landau cut is smoothened due to the folding. 

\section{Results and discussion}

\begin{figure}
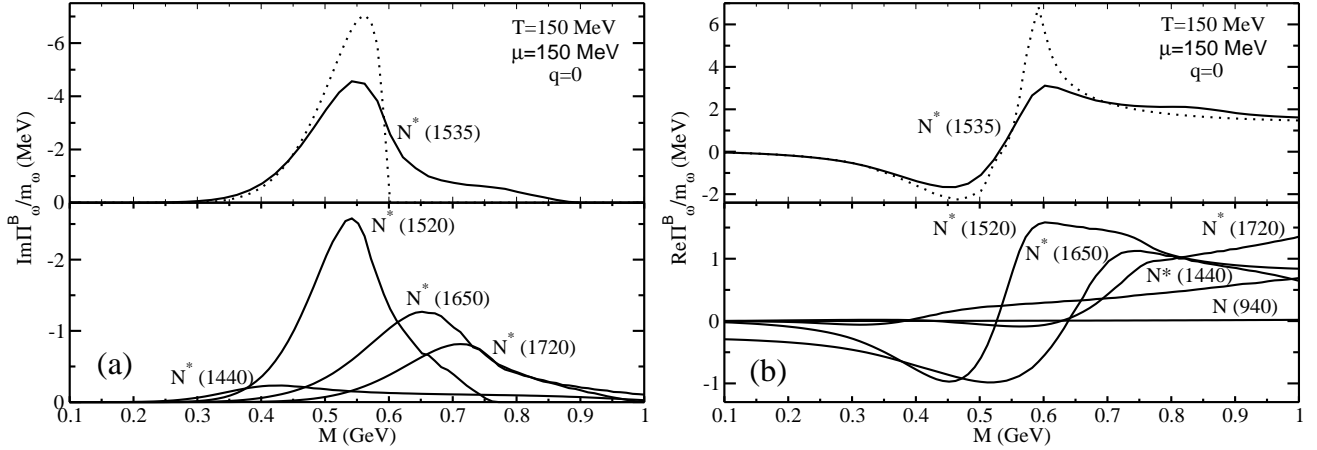

\begin{center}
\includegraphics[scale=0.35]{im_q0_B.eps}
\includegraphics[scale=0.35]{Re_q0_B.eps}
\end{center}
\caption{(a) The upper panel shows
the contribution of the $NN^*(1535)$ loop where the 
result in the narrow width approximation is indicated by the dotted line. The
lower panel shows the contribution of other $NB$ loops.
(b) The corresponding results for the real part.}
\label{B_q0}
\end{figure} 
We now present the results of numerical evaluation. We start with the
spin-averaged self-energy function defined as
\be
{\ov\Pi}=\frac{1}{3}(2{\ov\Pi}_{t}+q^2{\ov\Pi}_{l})
\label{pi_av}
\ee
where ${\ov\Pi}_{t,l}$ are defined in eq.~(\ref{pitpil}).
In Fig.~\ref{B_q0}(a) and (b) we plot the imaginary and real parts respectively of $\om$ 
self-energy for vanishing three-momentum. The contribution of the $NN^*(1535)$ loop is observed to 
play the most significant role
primarily due to the strong coupling of this resonance with the $\om N$ 
channel and is shown separately in
the upper panels. The effect of folding by the spectral function 
of the resonances denoted by $B$ is also shown
where the smoothing of the sharp cut-off in the imaginary part defining the
end of the Landau cut is clearly observed in the upper panel in (a). 
In the lower panels showing the 
contribution of the other loops the effect of the $N^*(1520)$ is seen to be
significantly more than the others. 
\begin{figure}
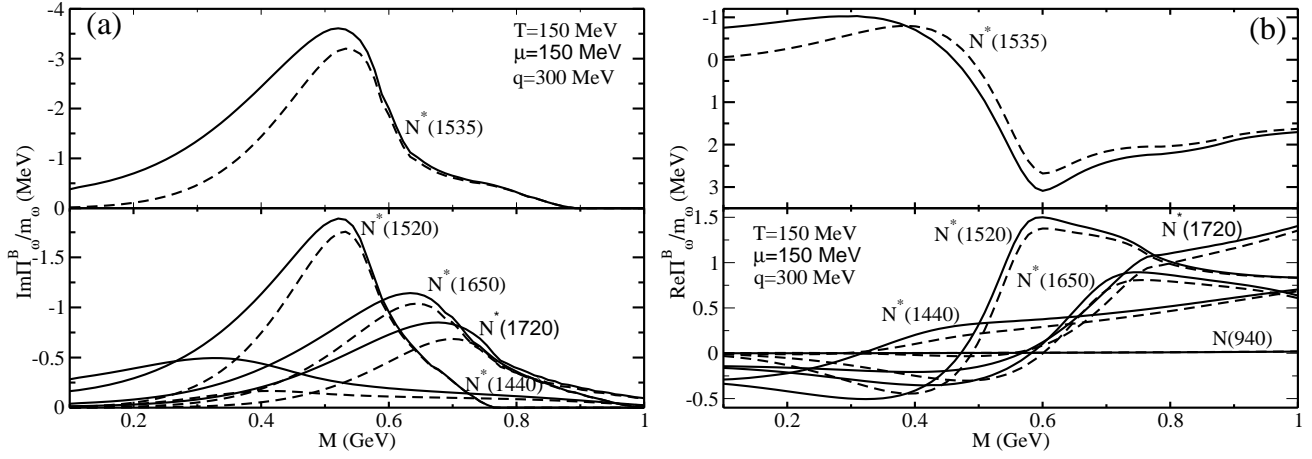

\begin{center}
\includegraphics[scale=0.35]{im_q300_B.eps}
\includegraphics[scale=0.35]{Re_q300_B.eps}
\end{center}
\caption{Imaginary (a) and real (b) parts of $\om$ self energy
for the $NB$ loops with three-momentum  $q=300$ MeV. Solid 
and dashed lines stand for transverse and longitudinal part of the self energy 
respectively.}
\label{B_q300}
\end{figure}  
In Fig.~\ref{B_q300}(a) and (b) we have shown the imaginary
and real parts respectively of the $\om$ self-energy for $\vq=300$ MeV. Here the transverse
component $\op_t$ is shown along with $q^2$ times the longitudinal component 
(note that $\op_t=q^2\op_l$ for $\vec q=0$).
As before, the $N^*(1535)$ makes the most important contribution and is
shown separately in the top panels. 

\begin{figure}
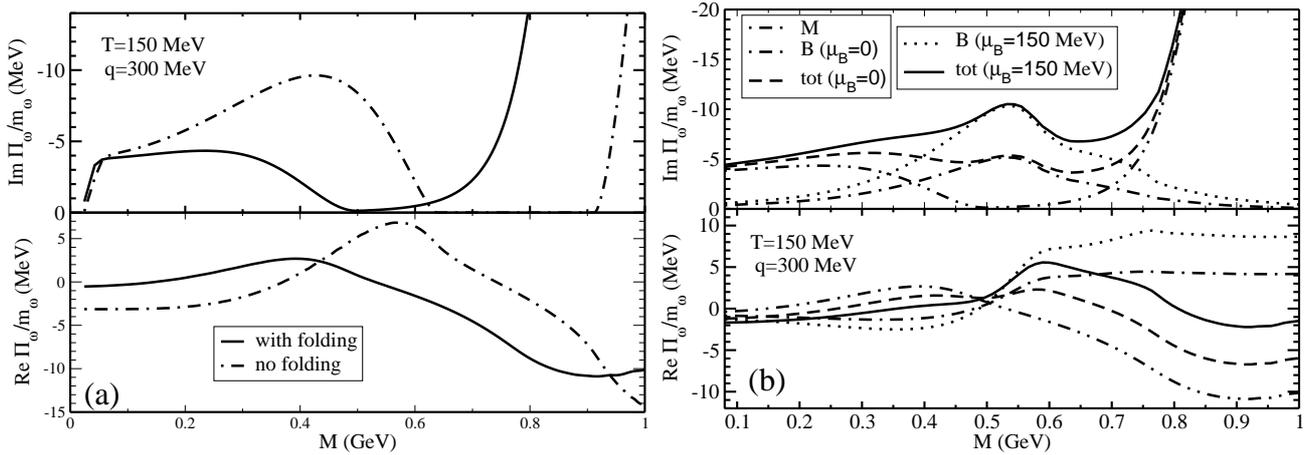

\begin{center}
\includegraphics[scale=0.35]{im_re_q300_M.eps}
\includegraphics[scale=0.35]{im_re_M_B_tot.eps}
\end{center}
\caption{(a) The imaginary (upper panel) and real (lower panel) part 
of $\om$ self energy for mesonic loop with and without the convolution with the
$\rho$ spectral function and (b) total imaginary (upper panel) and real (lower panel) 
parts of $\om$ self energy for meson and baryon loops.}
\label{M_Btot}
\end{figure} 
Plotted in Fig.~\ref{M_Btot}(a) is the spin averaged
$\om$ self-energy from the $\rho\pi$ loop. The effect of folding the $\rho\pi$
self-energy with the $\rho$ width is clearly visible in the upper panel by
the solid line which shows a finite contribution at the $\om$ pole instead
of a vanishing contribution in this region when this folding is not done, as
shown by the dashed line. 
This is because the $\om$ pole lies in between the Landau and unitary cut thresholds 
at $\sim$630 and $\sim$910 MeV respectively. In Fig.~\ref{M_Btot}(b) is shown the total
contribution from the meson and baryon loops for two values of the baryon
chemical potential. A noticeable contribution is seen in the imaginary part 
below the nominal $\om$ mass. In the lower panel is shown the real part 
where the meson and baryon loops provide a negative and positive contribution
respectively at the $\om$ pole which will be manifested in the spectral function.

\begin{figure}
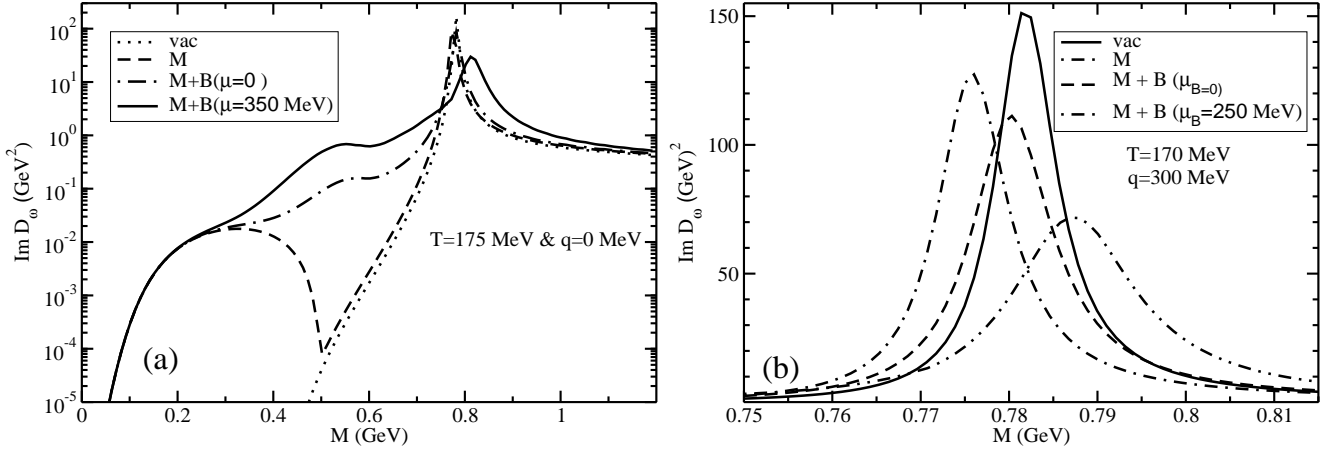

\begin{center}
\includegraphics[scale=0.35]{spec_log.eps}
\includegraphics[scale=0.35]{spec_M_B.eps}
\end{center}
\caption{
(a) The $\om$ spectral function showing individual contributions 
due to mesonic and baryonic loops and (b) the region close to the $\om$ pole.}
\label{mu0}
\end{figure} 
We now turn to the results for the spin
averaged spectral function of the $\om$  which is defined as 
\be
\mathrm{Im}\,\od(q)=\frac{1}{3}(2\mathrm{Im}\,\od_t+q^2\mathrm{Im}\,\od_l)
\label{spavgG}
\ee
where the transverse and longitudinal components are associated with the
corresponding projectors $P_\mn$ and $Q_\mn$ respectively.
These are given by
\be
\mathrm{Im}\,\od_{t,l}(q)=\frac{-\sum\iop_{t,l}^{B+M}}{(q^2-m_\om^2-
(1,q^2)\sum\mathrm{Re}\,\overline\Pi_{t,l}^{
B+M})^2+\{(1,q^2)\sum\iop_{t,l}^{B+M}\}^2}~.
\ee
It is interesting to note that the rate of lepton pair production in the late
(hadronic) stages of heavy ion collisions is essentially determined by the
spectral function of low mass vector mesons, $\rho$, $\omega$ and $\phi$.
This is given by (see e.g.~\cite{Gale}),
\be
\frac{dN}{d^4q
d^4x}=-\frac{\alpha^2}{3\pi^3q^2}f_{BE}(q_0)g^\mn\sum_{V=\rho,\omega,\phi}
{F^2_V}{m_V^2}\ {\rm Im}\overline D^V_\mn(q_0,\vec q)
\ee
where the exact propagator $\overline D_\mn(q_0,\vec q)$ for a vector meson 
$V$ is defined in eq.~(\ref{Dmn}). Using the relations
$g^\mn P_\mn=-2$ and $g^\mn Q_\mn=-q^2$ we get
${\rm Im}\overline D=-\frac{1}{3}g^\mn{\rm Im}\overline D_\mn$ which is the spin
averaged spectral function given by eq.~(\ref{spavgG}). We then have,
\be
\frac{dN}{d^4q
d^4x}=\frac{\alpha^2}{\pi^3q^2}f_{BE}(q_0)\sum_{V=\rho,\omega,\phi}
{F^2_V}{m_V^2}\ {\rm Im}\overline D^V(q_0,\vec q)~.
\ee

We now show
the contributions of the different loops to the spectral function.
To bring out the relative strengths at low invariant masses a logarithmic scale
is employed in Fig.~\ref{mu0}(a). The dashed line represents $\rho\pi$ loop 
in which the Landau cut 
contribution falls off in the vicinity of $M=m_\rho-m_\pi$ 
and then increases as the unitary cut contribution builds up.
The Landau cut contributions from the baryonic loops, 
shown by the solid and dash-dotted lines,
however dominate in the region below the $\om$ mass.
We now concentrate on a small $M$ range around the $\omega$ mass in  
Fig.~\ref{mu0}(b).
In tune with the real part of the self-energy shown in the lower panel of
Fig.~\ref{M_Btot}(b), the peak shifts a little to the left for the meson loop 
in contrast to the situation when baryonic contributions are added. The
slight increase in mass in this case is also accompanied by a larger 
imaginary part causing more suppression of the spectral strength at the peak.
\begin{figure}
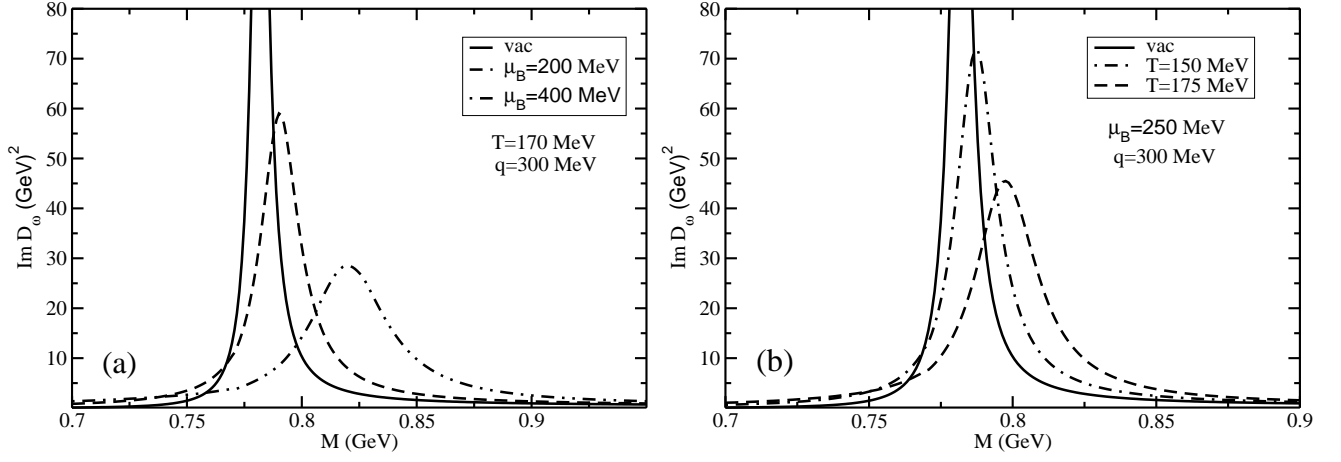

\begin{center}
\includegraphics[scale=0.35]{spec_mu.eps}
\includegraphics[scale=0.35]{spec_T.eps}
\end{center}
\caption{The spectral function of $\om$ for different values of (a) $\mu_B$  
and (b) $T$}
 \label{spec}
\end{figure} 
Next we plot the spectral function for different $\mu_{B}$ and $T$ in 
Fig.~\ref{spec}(a) and (b) respectively for $M$ close to the
$\om$ mass. As before, the small positive thermal mass shift of
the $\om$ increases with $\mu_{B}$ and $T$. The corresponding
decrease of the $\om$-spectral function at the peak representing the  
enhancement of width with increasing $\mu_{B}$ and $T$ is also seen.

\begin{figure}
\begin{center}
\includegraphics[scale=0.35]{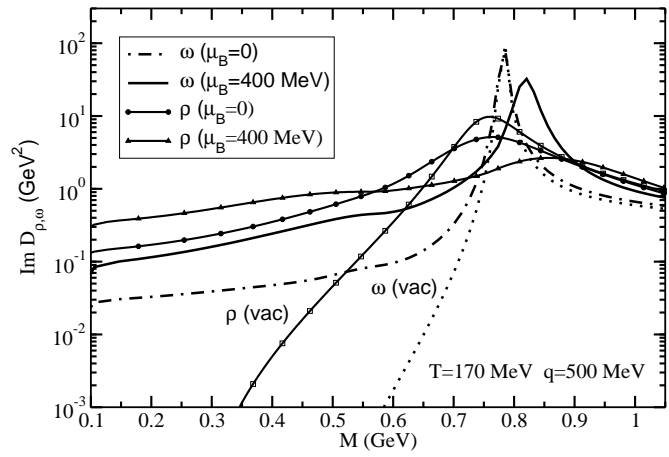}
\end{center}
\caption{The $\om$ spectral function seen in comparison with the
$\rho$}
\label{rho-om}
\end{figure}
In view of the fact that the $\rho$ and $\om$ peaks are close to each other it is
worthwhile to compare their relative spectral strengths below their nominal
masses. In Fig.~\ref{rho-om}(a) we have plotted the $\om$ spectral function at two values of the chemical
potential along with that of the $\rho$ which has been recently calculated in
Ref.~\cite{GS_NPA}. The sharp peak of the $\omega$ is stands out against the
smooth profile of the $\rho$. 
The characteristic $2\pi$ and $3\pi$ thresholds for the $\rho$ and $\omega$
in the vacuum case are also visible. 
Though the spectral strength of the $\om$
is lower than the $\rho$ they do have a sizeable
contribution in the region below $\sim$700 MeV.

\section{Summary and Discussion}

To summarize, we have evaluated the  spectral function of the $\om$ meson
in hot and dense matter using the framework of thermal field theory in the
real-time formulation. Using effective interactions,
one-loop self-energy graphs were evaluated for an extensive set of 
spin one-half and three-half $N^*$ resonances in addition to the $\rho\pi$ loop
using fully relativistic propagators and off-shell corrections for spin
three-half fields. The
imaginary part has been obtained from the discontinuities
 of the self-energy function which
provides an unified treatment of various scattering and decay processes 
occurring
in the thermal medium. Results of the real and imaginary parts
for all the loops and the full spectral function were presented
for several combinations of temperature, baryon chemical potential
and three-momenta relevant in heavy ion collisions.

We have made a comparison 
with the $\rho$ spectral function finding the $\om$ contribution to be
lower but of comparable magnitude. However, the fact that the latter is suppressed by a factor $\sim
10$ compared to the $\rho$ in the dilepton emission rate makes a quantitative
study of the $\om$ difficult. Additional hindrances could arise due to matter
induced $\rho-\omega$ mixing~\cite{Proy}. Nevertheless, the contribution of the
$\om$ spectral strength is essential for a quantitative description of the
dilepton data from heavy ion collisions~\cite{Sarkar_JPG,GSA_EPJC}. In view of high
quality data expected in future from heavy ion collisions at the FAIR facility
at GSI we can conclude that a detailed evaluation of the spectral strength
at finite temperature and baryon density is necessary for a quantitative 
analysis. 
%\newpage
\section{Appendix-A}
\setcounter{equation}{0}
\renewcommand{\theequation}{A.\arabic{equation}}

\begin{figure}
\centerline{\includegraphics[scale=0.7]{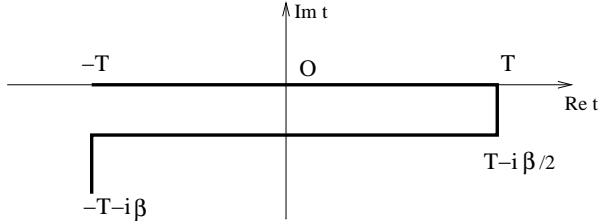}}
\caption{The contour $C$ in the complex time plane used here for the real
time formulation.}
\label{contour}
\end{figure}

In this appendix we review the real time formulation of equilibrium 
thermal field theory~\cite{Kobes,Bellac} leading to the
spectral representation of the vector boson propagator~\cite{Mallik_RT}. 
This formulation 
begins with a comparison between the time evolution operator
$e^{-iH(t_2-t_1)}$ of quantum theory and the Boltzmann weight $e^{-\bet H}
=e^{-iH(\tau-i\bet-\tau)}$ of statistical physics, where we introduce $\tau$ as a 
complex variable. Thus while for the time evolution operator, the times $t_1$ and
$t_2$ $~(t_2 > t_1)$ are any two points on the real line, the Boltzmann
weight involves a path from $\tau$ to $\tau-i\bet$ in the complex time
plane. Setting this $\tau =-T$, where $T$ is real, positive and large
($\to\infty$), we
get the symmetric contour $C$ shown in Fig.~\ref{contour}, lying within the region of analyticity
in this plane and accommodating real time correlation functions \cite{Mills,Niemi}.

Let the interacting (vector) bosonic field in the Heisenberg representation be
denoted by $V_\mu(x)$.
The thermal expectation value of the product $V_\mu(x)V^\dag_\nu(x')$ may
be expressed as
\be
\la V_\mu(x)V^\dag_{\nu}(x')\ra=\frac{1}{Z}\sum_{m,n}e^{-\beta E_m} \la
m|V_\mu(x)|n\ra\la n|V^\dag_{\nu}(x')|m\ra\,,
\label{expctvalue}
\ee
where $\la O\ra$ 
for any operator $O$ denotes the ensemble average;
\be
\la O\ra = {\rm Tr}(e^{-\beta H} O)/Z\,, ~~~~~~~~Z=Tre^{-\beta H}\,.
\ee
Note that we have two sums in (\ref{expctvalue}), one to evaluate the trace and the other
to separate the field operators. 
They run over a complete set of states,
which we choose as eigenstates $|m\ra$ of four-momentum $P_\mu$ with 
eigenvalues $(E_m,k_m)$. Using 
translational invariance of the field operator,
\be
V_\mu(x)=e^{iP\cdot x}V_\mu(0)e^{-iP\cdot x}\,,
\ee
we get
\be
\la V_\mu(x)V^\dag_{\nu}(x')\ra=\frac{1}{Z}\sum_{m,n}e^{-\beta E_m}\,e^{i(k_m-k_n)\cdot (x-x')}
\la m|V_\mu(0)|n\ra\la n|V^\dag_{\nu}(0)|m\ra\,.
\ee
Its spatial Fourier transform is
\bea
&&\negs \int d^3x\,e^{-i\vk\cdot(\vx-\vx')}\la V_\mu(x)V^\dag_{\nu}(x')\ra\no\\
&&\negs =\frac{(2\pi)^3}{Z} \sum_{m,n}e^{-\beta E_m}\,e^{i(E_m-E_n)(\tau-\tau')}
\de^3(\vk_m-\vk_n+\vk)\la m|V_\mu(0)|n\ra\la n|V^\dag_{\nu}(0)|m\ra\,,
\label{spatFT}
\eea
where the times $\tau,\, \tau'$ are on the contour $C$. We now insert unity on the 
left of eq.~(\ref{spatFT}) in the form
\[1=\int_{-\infty}^\infty dk_0' \de(E_m-E_n+k_0')\,.\]
Then it may be written as 
\be
\int d^3x \,e^{-i\vk\cdot(\vx-\vx')}\la V_\mu(x)V^\dag_{\nu}(x')\ra
=\int\frac{dk_0'}{2\pi}e^{-ik_0'(\tau-\tau')}M^+_{\mn}(k_0',\vk)\,,
\label{ft1}
\ee
where the spectral function $M^+_{\mn}$ is given by $[k'_\mu=(k_0',\vk)]$
\be
M^+_{\mn}(k')=\frac{(2\pi)^4}{Z}\sum_{m,n}e^{-\beta E_m}\,\de^4(k_m-k_n+k')
\la m|V_\mu(0)|n\ra\la n|V^\dag_{\nu}(0)|m\ra\,.
\label{mplus}
\ee

In the same way, we can work out the Fourier transform of 
$\la V^\dag_{\nu}(x')V_\mu(x)\ra$ 
\be
\int d^3x \,e^{-i\vk\cdot(\vx-\vx')}\la V^\dag_{\nu}(x')V_\mu(x)\ra
=\int\frac{dk_0'}{2\pi}e^{-ik_0'(\tau-\tau')}M^-_{\mn}(k_0',\vk)\,,
\label{ft2}
\ee
with a second spectral function $M^-_{\mn}$ given by
\be
M^-_{\mn}(k')=\frac{(2\pi)^4}{Z}\sum_{m,n}e^{-\beta E_m}\,\de^4(k_n-k_m+k')
\la m|V^\dag_{\nu}(0)|n\ra\la n|V_\mu(0)|m\ra\,.
\label{mminus}
\ee
The two spectral functions are related by the KMS relation \cite{Kubo,Martin}
\be
M^+_{\mn}(k)=e^{\beta k_0}M_{\mn}^-(k)\,,
\label{KMS}
\ee
in momentum space, which may be obtained by interchanging the dummy indices 
$m,n$ in one of $M^\pm_{\mn}(k)$ and using the energy conserving $\de$-function.

We next introduce the difference of the two spectral functions,
\be
\rho_{\mn}(k) \equiv M_{\mn}^+(k)-M_{\mn}^-(k)\,,
\label{diff}
\ee
and solve this identity and the KMS relation (\ref{KMS}) for $M^\pm_{\mn}(k)$, 
\be
M^+_{\mn} (k)=\{1+f(k_0)\}\rho_{\mn} (k)\,, ~~~~ M^-_{\mn}(k)=f(k_0)\rho_{\mn}(k)\,,
\label{KMS1}
\ee
where $f(k_0)$ is the distribution-like function
\be
f(k_0)=\frac{1}{e^{\beta k_0}-1}\,,~~~~~~~~-\infty <k_0 < \infty\,.
\ee
In terms of the true distribution function
\be 
n(|k_0|)=\frac{1}{e^{\beta |k_0|}-1}\,,
\ee
it may be expressed as
\bea
f(k_0)&=& f(k_0)\{\tht (k_0)+\tht(-k_0)\}\no\\
&=&  n\ep(k_0)-\tht (-k_0)\,.
\label{ff1}
\eea

With the above ingredients, we now build the spectral representations for
the thermal propagators. The time-ordered propagator is given by
\bea
-iD_{\mn}(x,x')&=&\la T_c V_\mu(x) V^\dag_{\nu}(x')\ra \no\\
&=&\tht_c(\tau-\tau')\la V_\mu(x) V^\dag_{\nu}(x')\ra+\tht_c(\tau'-\tau)
\la V^\dag_{\nu}(x') V_{\mu}(x)\ra\,.
\eea
Using eqs.~(\ref{ft1},\ref{ft2},\ref{KMS1}), we see that its spatial Fourier transform is given by
 \cite{Mills}
\be
D_{\mn}(\tau-\tau',\vk)=i\int_{-\infty}^\infty\frac{dk_0'}{2\pi}\rho_{\mn}(k_0',\vk)
e^{-ik_0'(\tau-\tau')}\{\tht_c(\tau-\tau')+f(k_0')\}\,,
\ee

As $T\rw \infty$, the contour of Fig.~\ref{contour} simplifies, reducing essentially to
two parallel lines, one the real axis and the other shifted by $-i\bet/2$,
points on which will be denoted respectively by subscripts 1 and $2$, so
that $\tau_1=t,\, \tau_2=t-i\bet/2$ \cite{Niemi}. The propagator then consists of 
four pieces, which may be put in the form of a $2\times 2$ matrix.
The contour ordered $\th's$ may now be converted to the usual time ordered
ones. If $\tau,\tau'$ are both on line $1$ (the real axis), the $\tau$ and
$t$ orderings coincide, $\th_c(\tau_1-\tau'_1)=\th(t-t')$. If they are on
two different lines, the $\tau$ ordering is definite,
$\th_c(\tau_1-\tau'_2)=0,\, \th_c(\tau_2-\tau'_1)=1$. Finally if they are
both on line $2$, the two orderings are opposite,
$\th_c(\tau_2-\tau'_2)=\th(t'-t)$.

Back to real time, we can work out the usual temporal Fourier transform of
the components of the matrix to get
\be   
\bD_{\mn} (k_0,\vk)=\int_{-\infty}^{\infty}\frac{dk_0'}{2\pi}\rho_{\mn}(k_0',\vk)
\bL (k_0',k_0)\,,  
\label{d11matrix}
\ee
where the elements of the matrix $\bL$ are given by \cite{Mallik_RT}
\bea
&& \Lm_{11}=-\Lm_{22}^* =\frac{1}{k_0'-k_0-i\eta}+2\pi if(k_0')\de(k_0'-k_0)\,,
\no\\
&& \Lm_{12}=\Lm_{21}=2\pi ie^{\beta k_0'/2}f(k_0')\de(k_0'-k_0)\,.
\label{Lm1}
\eea
Using relation (\ref{ff1}), we may rewrite (\ref{Lm1}) in terms of $n$, 
\bea
&& \Lm_{11}=-\Lm_{22}^* =\frac{1}{k_0'-k_0-i\eta\ep(k_0)}+2\pi in\ep(k_0)
\de(k_0'-k_0)\,,\no\\
&& \Lm_{12}=\Lm_{21}=2\pi i\sqrt{n(1+n)}\ep(k_0)\de(k_0'-k_0)\,.
\label{Lm2}
\eea

The matrix $\bL$ and hence the propagator $\bD_{\mn}$ can be diagonalised
to give 
\be
\bD_{\mn}(k_0,\vk)=\bU 
\left(\begin{array}{cc}\oD_{\mn} & 0\\0 &
-\oD_{\mn}^{\,*}\end{array}\right)\bU\,,
\label{diag}
\ee
where $\oD_{\mn}$ and $\bU$ are given by
\be
\oD_{\mn} (k_0,\vk)=\int_{-\infty}^{\infty}
\frac{dk_0'}{2\pi}\frac{\rho_{\mn}(k_0',\vk)}{k_0'-k_0-i\eta\ep(k_0)}\,,~~~~~~
\bU = \left( \begin{array}{cc} \sqrt{1+n} & \sqrt{n}\\\sqrt{n} &
\sqrt{1+n}\end{array}\right)\,.
\label{uuu}
\ee 
From here it is easily seen that
\be
\rho_{\mn}(k_0,\vk)=2\ep(k_0){\rm Im}\oD_{\mn}(k_0,\vk)~.
\ee
%Eq.~(\ref{diag}) shows that $\oD$ can be obtained from any of the elements of the
%matrix $\bD$, say $D_{11}$. Omitting the indices $\mn$, we get
%\be
%{\rm Re}\oD={\rm Re D}_{11}\,,~~~~{\rm Im}\oD=\tanh(\bet|k_0|/2){\rm Im D}_{11}\,.
%\label{tanh}
%\ee

Looking back at the spectral functions $M^{\pm}_{\mn}$  defined by
(\ref{mplus}, \ref{mminus}), we can 
express them as usual four-dimensional Fourier transforms of ensemble average of 
the operator products, so that $\rho_{\mn}$ is the Fourier transform of that of the 
commutator,
\be
\rho_{\mn}(k_0,\vk)=\int d^4y e^{ik\cdot (y-y')}\la
[V_\mu(y),V^\dag_{\nu}(y')]\ra\,,
\ee
where the time components of $y$ and $y'$ are on the real axis in the
$\tau$-plane. Taking the spectral function for the free scalar field,
\be
\rho_0=2\pi\ep(k_0)\de (k^2-m^2)\,,
\label{freesp}
\ee
we see that $\oD$ becomes the free propagator, $\oD (k_0,\vk)=-1/(k^2-m^2)$.
Eq.~(\ref{diag}) then yields the components of the free thermal
propagator given in eq.~(\ref{Dlam}).

\section{Appendix-B}
\setcounter{equation}{0}
\renewcommand{\theequation}{B.\arabic{equation}}

\subsection{Interaction Lagrangian and expressions for $L_{\mn}$ for baryonic loops}

The $\om N$ couplings with the resonances 
are described by the interaction Lagrangians~\cite{Muelich}    
\bea
{\cal L}&=&\frac{g_1}{2m_N}\ov{\psi}_B\sigma_{\mn}\del_\om^\nu\psi_N \om^{\mu} + h.c.
~~~~~~~~~~~~~~~~~~~~~~~~~~~~~~~~~~~~~~~~~~~~J^{P}_B=\frac{1}{2}^+ \nonumber\\
{\cal L}&=&-i\frac{g_1}{2m_N}\ov{\psi}_B \gamma^5\sigma_{\mn}\del_\om^\nu
 \psi_N \om^{\mu} + h.c.
~~~~~~~~~~~~~~~~~~~~~~~~~~~~~~~~~~~~~~~~~~~~J^{P}_B=\frac{1}{2}^- \nonumber\\
{\cal L}&=&-i[\ov{\psi}^{\mu}_B\gamma^5 (\frac{g_1}{2m_N}\gamma^{\alpha}+ i\frac{g_2}{4m^2_N}\del_N^{\alpha} +i\frac{g_3}{4m^2_N}\del_{\om}^{\alpha}) (\del^{\om}_{\alpha}{\cal O}_{\mn} - \del^{\om}_{\mu}{\cal O}_{\alpha\nu})\psi_N \om^{\nu} + h.c.]
~~~~~~J^{P}_B=\frac{3}{2}^+ \nonumber\\
{\cal L}&=&-[\ov{\psi}^{\mu}_B (\frac{g_1}{2m_N}\gamma^{\alpha}+ i\frac{g_2}{4m^2_N}\del_N^{\alpha} +i\frac{g_3}{4m^2_N}\del_{\om}^{\alpha}) (\del^{\om}_{\alpha}{\cal O}_{\mn} - \del^{\om}_{\mu}{\cal O}_{\alpha\nu})\psi_N \om^{\nu} + h.c.]
~~~~~~~~~~~J^{P}_B=\frac{3}{2}^-
\label{lag1}
\eea
where 
$\sg^{\mn}=\frac{i}{2}[\gm^\mu\gm^\nu-\gm^\nu\gm^\mu]$ 
and ${\cal O}_{\mn}=g_{\mn}-\frac{1}{4}\gm_\mu\gm_\nu$ is 
the off-shell projector contracted with the vertices
containing spin $3/2$ fields~\cite{Peccei} which
contributes only when it is 
off the mass shell.
The values of all the coupling constants 
in the $\om NB$ Lagrangian are taken from Ref.~(\cite{Muelich,Leu_58}) 
and are given in Table~1.
\begin{table}
\begin{center}
%\label{tab}
\begin{tabular}{|c|c|c|c|c|}
\hline
& & & & \\
$B$ & $J^P$ & $g_1$ & $g_2$ & $g_3$ \\
& & & & \\
\hline
& & & & \\
$N(940)$ & $\frac{1}{2}^+$ & -0.79  & & -  \\
& & & & \\
$N^*(1440)$ & $\frac{1}{2}^+$ & -4.35 &  & - \\
& & & & \\
$N^*(1520)$ & $\frac{3}{2}^-$ & 3.35 & 4.80 & -9.99  \\
& & & & \\
$N^*(1535)$ & $\frac{1}{2}^-$ & 6.50 &  & -  \\
& & & & \\
$N^*(1650)$ & $\frac{1}{2}^-$ & -3.27 &  & - \\
& & & & \\
$N^*(1720)$ & $\frac{3}{2}^+$ & -6.82 & -5.84 & -8.63 \\
& & & & \\
\hline
\end{tabular}
\caption{Table showing the coupling constants of $\om NB$ vertex where $B$
stands for various resonances considered.}
\end{center}
\end{table}

For spin $\frac{1}{2}^{\pm}$ resonances the tensor $L^{\mn}(k,q)$ of eq.~(\ref{Pimn-Lmn}) is given by
\be
L^{\mn}(k,q)=(\frac{g_1}{2m_N})^2\,tr[\sigma^{\mu
\alpha}q_{\alpha}(\ks+bm_N)\sigma^{\nu \alpha}q_{\alpha}(\ks +a\qs +m_B) ]
\ee
where $b=\pm1$ for $J^P=\frac{1}{2}^{\pm}$ resonances. On simplification it can
be put in the form 
\be
L^{\mn}=4(\frac{g_1}{2m_N})^2[(k^2-a(q\cdotp k)+bm_Nm_B)q^2A^{\mn}+2B^{\mn}+(0)C^{\mn}]
\ee
with
\bea
A_{\al\beta}(q)&=&-g_{\al\beta}+{q_\al q_\beta}/{q^2} ,\nonumber\\
B_{\al\beta}(k,q)&=&q^2 k_\al k_\beta-q\cdot k(q_\al k_\beta+k_\al q_\beta)
+(q\cdot k)^2g_{\al\beta} ,\nonumber\\
C_{\al\beta}(k,q)&=&q^4 k_\al k_\beta-q^2(q\cdot k)(q_\al k_\beta+k_\al
q_\beta) +(q\cdot k)^2q_\al q_\beta .
\eea
It is easily seen that these tensors vanish when contracted with $q^\alpha$.
Consequently, $L^{\mn}$ and hence the self-energy functions expressed in terms
of these tensors are all four-dimensionally transverse.

For spin $\frac{3}{2}^{\pm}$ resonances
\be
L^{\mn}(k,q)=-tr[V^{\mu\alpha}(\ks+bm_N)V^{\nu\beta}(\ks +a\qs +m_B)K_{\beta\alpha}]
\ee
where $V^{\mu\alpha} = V^{\mu\alpha}_0+c V^{\mu\alpha}_c$ for the off shell projection operator 
${\cal O}_{\mn}=g_{\mn}+c\gm_\mu\gm_\nu$ (i.e. $c=-\frac{1}{4}$) with
\bea
&&V^{\mu\alpha}_0=\frac{g_1}{2m_N}(\qs g^{\mu\alpha}-\gamma^{\mu}q^{\alpha}) +\frac{g_2}{4m_N^2}\{(q\cdotp k)g^{\mu\alpha}-k^{\mu}q^{\alpha}\} -\frac{g_3}{4m_N^2}(q^2g^{\mu\alpha}-q^{\mu}q^{\alpha})
\nonumber\\
&&V^{\mu\alpha}_c=\frac{g_1}{2m_N}\gamma^{\mu}(\gamma^{\alpha}\qs - 
\qs \gamma^{\alpha})+\frac{g_2}{4m_N^2}\gamma^{\mu}\{\gamma^{\alpha}(q\cdotp k)-\qs  
k^{\alpha}\}-\frac{g_3}{4m_N^2}\gamma^{\mu}(q^2\gamma^{\alpha} -\qs
q^{\alpha})~. 
\eea
Here $b$ indicates the parity states of spin $3/2$ resonances i.e. $b=\pm1$ 
for $J^P=\frac{3}{2}^{\pm}$.
In this case $L^{\mn}$ can be expressed as 
%tensor for spin $3/2$ resonances can be expressed as
\be
L^{\mn}=(\frac{g_1}{2m_N})^2L^{\mn}_{11}+(\frac{g_2}{4m_N^2})^2L^{\mn}_{22}+(\frac{g_3}{4m_N^2})^2L^{\mn}_{33}+\frac{g_1}{2m_N}\frac{g_2}{4m_N^2}L^{\mn}_{12}+\frac{g_1}{2m_N}\frac{g_3}{4m_N^2}L^{\mn}_{13}+\frac{g_2}{4m_N^2}\frac{g_3}{4m_N^2}L^{\mn}_{23}
\ee
where
\be
L^{\mn}_{ij}=(\alpha^{00}_{ij}+c\alpha^{0c}_{ij}+c^2\alpha^{cc}_{ij})A^{\mn}+(\beta^{00}_{ij}+c\beta^{0c}_{ij}+c^2\beta^{cc}_{ij})B^{\mn}+(\gamma^{00}_{ij}+c\gamma^{0c}_{ij}+c^2\gamma^{cc}_{ij})C^{\mn}
\ee
with six possible sets of $ij$ ($ij=11,22,33,12,13\ {\rm and}\ 23$). The
structure of the $L^{\mn}_{ij}$ ensures that the self-energies are explicitly transverse.
%and those coefficients of $A^{\mn}$, $B^{\mn}$, $C^{\mn}$
The values of the coefficients for each set are given by
\bea
&&\alpha^{00}_{11}=\frac{8}{3m_B^2}[(k^2m_B^2+bm_Nm_B^3-k^2q^2)-a(q\cdotp k)(2k^2+q^2+2a(q\cdotp k))]q^2
\nonumber\\
&&\beta^{00}_{11}=\frac{8}{3m_B^2}[k^2+m_B^2+a(q\cdotp k)]
\nonumber\\
&&\alpha^{0c}_{11}=4\frac{8}{3m_B^2}[bm_Nm_BS_N^2-a(q\cdotp k)(S_N^2+3a(q\cdotp k)-2bm_Nm_B)]q^2
\nonumber\\
&&\beta^{0c}_{11}=4\frac{8}{3m_B^2}[k^2-m_B^2+2a(q\cdotp k)]
\nonumber\\
&&\gamma^{0c}_{11}=4\frac{8}{3m_B^2}
\nonumber\\
&&\alpha^{cc}_{11}=4\frac{8}{3m_B^2}[(m_N^2+2bm_Nm_B)\{S_N^2+2a(q\cdotp k)\}-a(q\cdotp k)\{S_N^2+4a(q\cdotp k)\}]q^2
\nonumber\\
&&\beta^{cc}_{11}=4\frac{8}{3m_B^2}[2\{k^2-m_B^2+2a(q\cdotp k)\}]
\nonumber\\
&&\gamma^{cc}_{11}=4\frac{8}{3m_B^2}[2]
\nonumber\\
\eea
\bea
&&\beta^{00}_{22}=\frac{8}{3m_B^2}[\{k^2-bm_Nm_B+a(q\cdotp k)\}m_B^2]
\nonumber\\
&&\gamma^{00}_{22}=\frac{8}{3m_B^2}[-k^2+bm_Nm_B-a(q\cdotp k)]
\nonumber\\
&&\beta^{0c}_{22}=\frac{8}{3m_B^2}[-bm_Nm_B\{S_N^2+2a(q\cdotp k)\}]
\nonumber\\
&&\beta^{cc}_{22}=\frac{8}{3m_B^2}[a(q\cdotp k)\{S_N^2+2a(q\cdotp k)\}]
\nonumber\\
&&\gamma^{cc}_{22}=\frac{8}{3m_B^2}[(k^2-bm_Nm_B)\{S_N^2+2a(q\cdotp k)\}]
\nonumber\\
\eea
\bea
&&\alpha^{00}_{33}=\frac{8}{3m_B^2}[\{-k^2+bm_Nm_B-a(q\cdotp k)\}m_B^2]q^4
\nonumber\\
&&\gamma^{00}_{33}=\frac{8}{3m_B^2}[-k^2+bm_Nm_B-a(q\cdotp k)]
\nonumber\\
&&\alpha^{0c}_{33}=\frac{8}{3m_B^2}[bm_Nm_B\{S_N^2+2a(q\cdotp k)\}]q^4
\nonumber\\
&&\gamma^{0c}_{33}=\frac{8}{3m_B^2}[-2\{S_N^2+2a(q\cdotp k)\}]
\nonumber\\
&&\alpha^{cc}_{33}=\frac{8}{3m_B^2}[\{k^2-2bm_Nm_B+a(q\cdotp k)\}\{S_N^2+2a(q\cdotp k)\}]q^4
\nonumber\\
&&\gamma^{cc}_{33}=\frac{8}{3m_B^2}[-2\{S_N^2+2a(q\cdotp k)\}]
\nonumber\\
\eea
\bea
&&\alpha^{00}_{12}=\frac{8}{3m_B^2}[-k^2+2bm_Nm_B-a(q\cdotp k)]m_B (q\cdotp k)q^2
\nonumber\\
&&\beta^{00}_{12}=\frac{8}{3m_B^2}[k^2-2bm_Nm_B+a(q\cdotp k)+m_B^2]m_B
\nonumber\\
&&\gamma^{00}_{12}=\frac{8}{3m_B^2}[-m_B]
\nonumber\\
&&\alpha^{0c}_{12}=2\frac{8}{3m_B^2}[abm_N(q\cdotp k)\{S_N^2+2a(q\cdotp k)\}]q^2
\nonumber\\
&&\beta^{0c}_{12}=2\frac{8}{3m_B^2}[(2m_B-bm_N)\{S_N^2+2a(q\cdotp k)\}]
\nonumber\\
&&\alpha^{cc}_{12}=4\frac{8}{3m_B^2}[abm_N(q\cdotp k)\{S_N^2+2a(q\cdotp k)\}]q^2
\nonumber\\
&&\beta^{cc}_{12}=4\frac{8}{3m_B^2}[(2m_B-bm_N)\{S_N^2+2a(q\cdotp k)\}]
\nonumber\\
\eea
\bea
&&\alpha^{00}_{13}=\frac{8}{3m_B^2}[(k^2+q^2+m_B^2-2bm_Nm_B)(q\cdotp k)+aq^2(k^2-2bm_Nm_B)]m_Bq^2
\nonumber\\
&&\beta^{00}_{13}=\frac{8}{3m_B^2}[-a m_B q^2]
\nonumber\\
&&\alpha^{0c}_{13}=2\frac{8}{3m_B^2}[\{(2m_B-bm_N)(q\cdotp k)-abm_Nq^2\}\{S_N^2+2a(q\cdotp k)\}]q^2
\nonumber\\
&&\alpha^{cc}_{13}=4\frac{8}{3m_B^2}[\{(2m_B-bm_N)(q\cdotp k)-abm_Nq^2\}\{S_N^2+2a(q\cdotp k)\}]q^2
\nonumber\\
\eea
\bea
&&\alpha^{00}_{23}=2\frac{8}{3m_B^2}[k^2-2bm_Nm_B+a(q\cdotp k)]m_B^2(q\cdotp k)q^2
\nonumber\\
&&\gamma^{00}_{23}=2\frac{8}{3m_B^2}[-a\{k^2-2bm_Nm_B+a(q\cdotp k)\}]
\nonumber\\
&&\alpha^{0c}_{23}=2\frac{8}{3m_B^2}[-bm_Nm_B(q\cdotp k)\{S_N^2+2a(q\cdotp k)\}]q^2
\nonumber\\
&&\gamma^{0c}_{23}=2\frac{8}{3m_B^2}[-a\{S_N^2+2a(q\cdotp k)\}]
\nonumber\\
&&\alpha^{cc}_{23}=2\frac{8}{3m_B^2}[(k^2-2bm_Nm_B)(q\cdotp k)\{S_N^2+2a(q\cdotp k)\}]q^2
\nonumber\\
&&\beta^{cc}_{23}=2\frac{8}{3m_B^2}[-a\{S_N^2+2a(q\cdotp k)\}]q^2
\eea
The rest of the coefficients are zero.
%($\gamma^{00}_{11}$, $\alpha^{00}_{22}$, $\alpha^{0c}_{22}$, $\alpha^{cc}_{22}$, $\gamma^{0c}_{22}$, $\beta^{00}_{33}$,
%$\beta^{0c}_{33}$, $\beta^{cc}_{33}$, $\gamma^{0c}_{12}$, $\gamma^{cc}_{12}$, $\gamma^{00}_{13}$,
%$\beta^{0c}_{13}$, $\beta^{cc}_{13}$, $\gamma^{0c}_{13}$, $\gamma^{cc}_{13}$, $\beta^{00}_{23}$, $\beta^{0c}_{23}$
%and $\gamma^{cc}_{23}$)
%are appeared to be zero. 

\subsection{Interaction Lagrangian and expressions for $L_{\mn}$ for mesonic loops}
For the interaction vertices entering the self-energy graphs for mesonic loops, 
we expand the relevant terms of the chiral Lagrangian and retain the lowest order 
terms to get~\cite{Ecker,MS2}
\be
{\cal L}_{int}=\frac{g_{m}}{\F}\ep_{\mu\nu\lm\sg}(\del^\nu\om^\mu\vec\rho^\lm-
\om^\mu\del^\nu\vec\rho^\lm)\cdot\del^\sg\vec\pi~.
\ee
Here, the pion decay constant, $\F=93$ MeV. The 
decay rate $\Gm (\omega \to 3\pi) = 7.6$ MeV gives
$g_{m}=5.5$.

The expression for $L^{\mn}$ 
appearing in the $\om$ self-energy
for the $\pi\rho$ loop is given by,
\be
L^{\mn}_{(\rho\pi)}(q,k)=-4\left(\frac{g_m}{\F}\right)^2(B^{\mn}+q^2k^2
A^{\mn})
\ee
and is thus explicitly transverse.

\section{Appendix-C}
\setcounter{equation}{0}
\renewcommand{\theequation}{C.\arabic{equation}}

Here we show how the real part of the self-energy obtained from a direct
evaluation of the diagrams can be put in a dispersion integral form.

Consider the case of a one-loop diagram with bosonic internal lines
\be 
\Pi^{11}(q)=i \int\frac{d^4k}{(2\pi^4)}D^{11}(k)D^{11}(q-k)
\label{self_11}
\ee
where we take constant vertices for the sake of illustration. 
Inclusion of momentum dependent vertices may need subtractions. 
Here the 11 component of the scalar
propagator matrix is
\be 
D_{11}(p)=\frac{-1}{p^2-m^2+i\eta}+2\pi in(\om)\delta(p^2-m^2)
~~,~~\om=\sqrt{\vp^2+m^2}~.
\ee
On multiplying out the propagators in (\ref{self_11}), we get three types
of terms,
\be 
\Pi^{11}(q)=\Pi^{11}_{(0)} + \Pi^{11}_{(1)} + \Pi^{11}_{(2)}
\ee
where $\Pi^{11}_{(0)}$, $\Pi^{11}_{(1)}$ and $\Pi^{11}_{(2)}$ are
terms without the Bose distribution $n(\om)$ (the vacuum term), and those linear 
and quadratic in $n(\om)$ respectively. In each of these we carry out the $k_0$ 
integration getting,
\bea 
\Pi^{11}_{(0)}&=& i\int\frac{d^4k}{(2\pi)^4}
\frac{1}{(k^2-m_1^2+i\eta)\{(q-k)^2-m_2^2+i\eta\}}
\nn\\
&=&\int \frac{d^3k}{(2\pi)^34\om_1\om_2}
(\frac{1}{q_0-\om_1-\om_2+i\eta}-\frac{1}{q_0+\om_1+\om_2-i\eta})~,
\eea
\bea 
\Pi^{11}_{(1)}&=& 2\pi\int\frac{d^4k}{(2\pi)^4}
[\frac{n_1\delta(k^2-m_1^2)}{(q-k)^2-m_2^2+i\eta} 
+ \frac{n_2\delta\{(q-k)^2-m_2^2\}}{k^2-m_1^2+i\eta}]
\nn\\
&=&\int \frac{d^3k}{(2\pi)^34\om_1\om_2}
[(n_1+n_2)(\frac{1}{q_0-\om_1-\om_2+i\eta}-\frac{1}{q_0+\om_1+\om_2-i\eta})
\nn\\
&&~~~~~~~~~~~~~~~~~+n_1(\frac{1}{q_0+\om_1-\om_2+i\eta}-\frac{1}{q_0-\om_1+\om_2-i\eta})
\nn\\
&&~~~~~~~~~~~+n_2(\frac{1}{q_0-\om_1+\om_2+i\eta}-\frac{1}{q_0+\om_1-\om_2-i\eta})]~,
\eea
\bea
\Pi^{11}_{(2)}&=&-i(2\pi)^2\int\frac{d^4k}{(2\pi)^4}
n_1n_2\delta(k^2-m_1^2)\delta\{(q-k)^2-m_2^2\}
\nn\\
&=&-2\pi i\int\frac{d^3k}{(2\pi)^3}\frac{n_1n_2}{4\om_1\om_2}
[\delta(q_0-\om_1-\om_2)+\delta(q_0-\om_1+\om_2)]
\nn\\
&&~~~~~~~~~~+\delta(q_0+\om_1-\om_2)+\delta(q_0+\om_1+\om_2)]~.
\eea
Note that $\Pi^{11}_{(2)}$ is completely imaginary. Adding the three pieces,
we get the complete self-energy integral as
\bea 
\Pi^{11}&=&\int \frac{d^3k}{(2\pi)^34\om_1\om_2}
[(1+n_1+n_2)(\frac{1}{q_0-\om_1-\om_2+i\eta}-\frac{1}{q_0+\om_1+\om_2-i\eta})
\nn\\
&&+n_1(\frac{1}{q_0+\om_1-\om_2+i\eta}-\frac{1}{q_0-\om_1+\om_2-i\eta})\nn\\
&&+n_2(\frac{1}{q_0-\om_1+\om_2+i\eta}-\frac{1}{q_0+\om_1-\om_2-i\eta})\nn\\
&&-2\pi i n_1n_2\{\delta(q_0-\om_1-\om_2)+\delta(q_0-\om_1+\om_2)
+\delta(q_0+\om_1-\om_2)+\delta(q_0+\om_1+\om_2)\}]~.
\label{pi_11}
\eea
We can easily separate it into real and imaginary parts.
The imaginary part is given by
\bea
{\rm Im}\Pi^{11}(q)&=&-\pi\int \frac{d^3k}{(2\pi)^34\om_1\om_2}[(1+n_1+n_2+2n_1n_2)
\{\delta(q_0-\om_1-\om_2)+\delta(q_0+\om_1+\om_2)\}
\nn\\
&&~~~~~~~~+(n_1+n_2+2n_1n_2)\{\delta(q_0-\om_1+\om_2)+\delta(q_0+\om_1-\om_2)\}]
\eea
Using
\bea
1+n_1+n_2+2n_1n_2&=&{\rm coth}(\beta q_0/2)(1+n_1+n_2)~,
\nn\\
n_1+n_2+2n_1n_2&=&-{\rm coth}(\beta q_0/2)(n_1-n_2)
\eea
\bea
{\rm Re}{\ov\Pi}&=&{\rm Re}\Pi^{11}
\nn\\
{\rm Im}{\ov\Pi}&=&\ep(q_0){\rm tanh}(\beta q_0/2){\rm Im}\Pi^{11}
\eea
and going over to the diagonalized form
\bea
{\rm Im}{\ov\Pi}(q)&=&-\pi\ep(q_0)\int \frac{d^3k}{(2\pi)^34\om_1\om_2}[(1+n_1+n_2)
\{\delta(q_0-\om_1-\om_2)-\delta(q_0+\om_1+\om_2)\}
\nn\\
&&~~~~~~~~-(n_1-n_2)\{\delta(q_0-\om_1+\om_2)-\delta(q_0+\om_1-\om_2)\}]~.
\label{im_pibar}
\eea
The real part may be obtained from (\ref{pi_11}) as
\bea
&&{\rm Re}{\ov\Pi}={\rm Re}\Pi^{11}=\int \frac{d^3k}{(2\pi)^34\om_1\om_2}
\left[(1+n_1+n_2){\cal
P}\left\{\frac{1}{q_0-\om_1-\om_2}-\frac{1}{q_0+\om_1+\om_2}\right\}\right.
\nn\\
&&~~~~~~~~~~~~~~~~~~~~~~~-(n_1-n_2)\left.{\cal P}\left\{\frac{1}{q_0-\om_1+\om_2}
-\frac{1}{q_0+\om_1-\om_2}\right\}\right]~.
\label{re_pibar}
\eea
Putting (\ref{im_pibar}) and (\ref{re_pibar}) together the complete ${\ov\Pi}$ 
can be written as
\bea 
{\ov\Pi}&=&\int \frac{d^3k}{(2\pi)^34\om_1\om_2}
\left[(1+n_1+n_2)\left(\frac{1}{q_0-\om_1-\om_2+i\eta\ep(q_0)}-\frac{1}{q_0+\om_1+\om_2+i\eta\ep(q_0)}\right)\right.
\nn\\
&&-(n_1-n_2)\left.\left(\frac{1}{q_0-\om_1+\om_2+i\eta\ep(q_0)}-\frac{1}{q_0+\om_1-\om_2+i\eta\ep(q_0)}\right)\right]
\label{pibar}
\eea
Note that Eq.~(\ref{MM_rho}) reduces to this form on putting the vertex
factors $L^\mn$ to unity.

At this point we recall that retarded (and advanced) functions have the right 
analytical properties to appear in the dispersion integral~\cite{Fetter}.
The imaginary part in the retarded continuation can be obtained from 
the diagonal element (with the bar) using the relation~\cite{Bellac}
\be
{\rm Im}\Pi(q_0,\vq)=\ep(q_0){\rm Im}\ov\Pi(q_0,\vq)~.
\label{im_piret}
\ee
Note that for $q_0>0$ as is the case here, the barred quantities are
numerically same as the retarded ones. 

It is now simple to convert Eq.~(\ref{re_pibar}) into a dispersion integral form by
introducing the decomposition of unity in the form
\be 
1=\int_{-\infty}^{\infty} dq'_0\delta(q'_0\mp \om_1\mp \om_2)
\ee
\bea
&&{\rm Re}{\Pi}=\int \frac{d^3k}{(2\pi)^34\om_1\om_2}{\cal P}
\int_{-\infty}^{\infty}\frac{dq'_0}{q_0-q'_0}
[(1+n_1+n_2)
\{\delta(q'_0-\om_1-\om_2)-\delta(q'_0+\om_1+\om_2)\}
\nn\\
&&~~~~~~~~-(n_1-n_2)\{\delta(q'_0-\om_1+\om_2)-\delta(q'_0+\om_1-\om_2)\}]
\eea
Interchanging the order of integrals, we get
\bea
&&{\rm Re}{\Pi}=\frac{1}{\pi}{\cal P}\int_{-\infty}^{\infty}\frac{dq'_0}{q'_0-q_0}[-\pi\int
 \frac{d^3k}{(2\pi)^34\om_1\om_2}
[(1+n_1+n_2)
\{\delta(q'_0-\om_1-\om_2)-\delta(q'_0+\om_1+\om_2)\}
\nn\\
&&~~~~~~~~-(n_1-n_2)\{\delta(q'_0-\om_1+\om_2)-\delta(q'_0+\om_1-\om_2)\}]]
\eea
Identifying the imaginary part from (\ref{im_pibar}) using (\ref{im_piret}),
\be
{\rm Re}{\Pi}(q_0,\vq)=\frac{1}{\pi}{\cal P}\int\frac{dq'_0}{q'_0-q_0}{\rm
Im}\Pi(q'_0,\vq)
\ee

In the case of an unstable particle in the loop, the width is introduced by
replacing its propagator by one which now contains an integral over the mass
weighted by its spectral function. We thus replace e.g.
$D_{11}(q-k,m_2)$ by $\int \rho(M) D_{11}(q-k,M)$ where $\rho(M)$ is the spectral
function~\cite{Mallik_RT}. The original self-energy is thus replaced by one which is a weighted 
sum over different slices of mass of the particle. The real and imaginary parts
hence follow the same relation as before.

\end{document}